\documentclass[a4paper,11pt]{article}
\usepackage{jheppub_kim}
\usepackage{pdflscape}
\usepackage{amsmath}
\usepackage{amssymb}
\usepackage{dcolumn}
\usepackage{bm}
\usepackage{color}
\usepackage{epsfig}
\usepackage{amsfonts}
\usepackage{graphicx}
\usepackage{subfigure}
\usepackage{dcolumn}
\begin{document}
\title{Quark anti-quark  entropic  force and  diffusion constant in a Yang-Mills like theory}
\author[]{S. Tahery}
\author[]{ and J. Sadeghi}
\affiliation{Sciences Faculty, Department of Physics, University of Mazandaran, 47416-95447, Babolsar, Iran}
\emailAdd {s.tahery@stu.umz.ac.ir}
\emailAdd {pouriya@ipm.ir}
\abstract{The entropic force  experienced by a moving quarkonia in a plasma, is computed in a hyperscaling violation   background by holography approach. In that case, the axis of the moving quarkonia has an arbitrary angle with wind. We use the results  for  HSV parameters which obtained in our paper \cite{hsvinve} as appropriate values satisfying condition $ImV_{Q\bar{Q}}<0$. We will show that, in this background the entropic force of the meson shows a decreasing behaviour corresponds to increasing dissociation length. Results of this paper are in agreement with our mentioned work. In addition we will find  diffusion constant in an exact case and by an  exact approach \cite{hypsh} which satisfies the mentioned condition in there. }
\keywords{entropic force, shear diffusion, moving quarkonia in a plasma, hyperscaling violation, dissociation of a meson}
\maketitle
\section{Introduction}
Studying of $Q\bar{Q}$ interaction needs the effect of the medium in the motion of the pair to be considered. Because this pair is not produced at rest in strongly coupled quark gluon plasma (QGP), the velocity of the pair through the plasma has some effects on its interactions that should be taken into account. The interaction energy has a finite imaginary part at finite temperature that can be used to estimate the thermal width of the quarkonia \cite{nbma,ybm}. Calculations of $ImV_{Q\bar{Q}}$ relevant to
QCD and heavy ion collisions were performed for static $Q\bar{Q}$ pairs using pQCD \cite{mlop} and
lattice QCD \cite{arth,gaca,gcs} before AdS/CFT.   Melting of heavy quarkonium is one of the main experimental signatures of the formation of QGP \cite{tmh}. The AdS/CFT is a correspondence \cite{jmm, ssg, ew, oas} between a string theory in AdS space and a conformal field theory in physical space-time.This theory describes the phenomenology of hadronic properties and demonstrate their ability to incorporate such essential properties
of QCD as confinement and chiral symmetry breaking. The study of the
moving heavy quarkonia in space-time with AdS/QCD approach plays important role in
interaction energy \cite{avi, mst, msd, mmk}.  One of the most important works was done by J. Noronha et al where they carried out the imaginary potential for $N = 4$ SYM theory in \cite{twlh}.  One another important quantity in study of $Q\bar{Q}$ is entropic force which is responsible to dissociate the quarkonium, because according to it's definition entropic force is a force which drives the system towards the state with a larger entropy \cite{epr}, in addition the entropic force may be responsible for gravity \cite{ogln} that we will not discuss here and interested reader can refer to the approriate reference. So, growing of the entropy $S$  with the inter-quark distance $L$ gives  the entropic force as we will study in this paper. Dissociation of the quarkonium is related to its entropic force\cite{desd}, it has been argued in lattice QCD studies such as \cite{hqa,okfz,ppkp}. \\
Here we discuss about gravity dual of QCD. Can we have an exact gravity dual of QCD? One can close to it to some extent \cite{astp}. What people do, is to start by considering $N=4$ SYM theory, which at long distances the theory reduces to pure Yang-Mills theory. But there are some differences between this YM theory and the gauge theory dual to the original  $N=4$ theory. Although the vacua of QCD and $N=4$ SYM theory have very different properties $N=4$ SYM at $T\neq 0$ with QCD $T> T_{c}$ (temperature $T_{c}$ of the crossover from a hadron gas to quark-gluon plasma), many of the qualitative distinctions disappear or become unimportant.\\
It is interesting to study hydrodynamic behaviour of a meson by considering a meson in a HSV background. In a wide variety of strongly coupled quantum field theories, viscosity makes a bound \cite{visst} which has been studied by holography \cite{largn}. In many of these theories, the shear viscosity $\xi$ satisfies $\frac{\xi}{s}=\frac{1}{4 \pi}$. In \cite{hypsh} the authors mentioned that from approach of \cite{hohy} one can study diffusion phenomenon with universal behaviour of the diffusion constant. They found exact relation for HSV parameters by holography approach . Proceeding by considering special  cases they found exact condition on $d$, $\theta$ and $z$ parameters which can be used to have diffusion constant.\\
The investigation of moving heavy quarkonia in a space-time  plays an
important role in the interaction energy \cite{mst,msd,mmk,gac} by AdS/QCD approach. The non-relativistic bound states in a moving thermal bath has been studied by \cite{nbs}. On the other hand   the heavy Quarkonium moving in a Quark-Gluon Plasma is another interesting work  which has been studied by \cite{drst, entrothermal}. Here we note that the different metric backgrounds lead us to face various effects of interaction energy.
The evaluation of Im$V_{Q\bar{Q}}$ will yield to determine the suppression of ${Q\bar{Q}}$ in heavy ion collision \cite{sif}. On the other hand the effects of deformation parameter in the thermal width of moving quarkonia in plasma has been studied in \cite{tec}.  Also the imaginary part of the static potential has been studied  as a observable and moving quarkonia  in strongly coupled anisotropic plasma \cite{ipap,osca,ihqm}. Here, in order to have Re $V_{Q\bar{Q}}$ and Im $V_{Q\bar{Q}}$  for ${Q\bar{Q}}$ in a plasma we use  boosted frame \cite{fn}.
From bottom-up  and top-down approaches and some gauge gravity dualities one can study the behaviour of some parameters in the corresponding theory. By bottom-up approach we consider different metric backgrounds and  study the behaviour of some parameters  in the thermal width of a moving quarkonia in plasma. One interesting case is considering metrics which are dual to the field theories and  are scale invariant but not conformally invariant. As we know such corresponding metrics have a Lifshitz  (with hyperscaling violation) scaling symmetries at quantum critical points and a dynamical critical exponent called $z$. We note that  the behaviour of the system with  hyperscaling violation metric backgrounds near to  phase transition, is characterized by the critical exponent $z$. Also we know that the  time and space in such background will be scaled differently. So the corresponding  metric is not invariant under the mentioned scaling. Here $d-\theta$ plays the role of an effective  dimensional space in the dual field theory  \cite{pd,sr},
and  theories with hyperscaling violation are intrinsically non-relativistic.
 With all above explanations, we organize the paper
as follows.\\
 In section 2 , we will review hyperscaling violation metric background and results of our last paper. Then we  apply the  case with relevant parameter  which we found in our paper \cite{hsvinve} in HSV metric to find entropic force for a moving dipole in section 3, so we can have a comparison between the results of two papers about meson dissociation.\\
 In section 4 we will calculate diffusion constant in our candidate HSV case. Section 5 is about our conclusion and results.

\section {Review of thermal width in a  hyperscaling violation metric background}
As we mentioned before, this paper is in continue of \cite{hsvinve}, so it is highly recomended to have a review on that. But since in current work we just apply that result to find entropic force in an exact case of hyperscaling violation background, who are not interested in thermal width, can skip from the mentioned paper and consider the review on what we have done as follows.
\subsection{Review on hyperscaling violation background} 
 In this section  we  review   hyperscaling violation (HSV) metric background.  Such a metric background is scale invariant but not conformally invariant. we consider the following metric,
\begin{equation}\label{metric1}
ds^2=-\frac{1}{r^{2z}} dt^2+ \frac{1}{r^2}(dr^2+dx_{i}^{2}),
\end{equation}
which is invariant under scaling
 $t\longrightarrow\lambda^{z} t $ , \quad $x_{i}\longrightarrow\lambda x_{i} $, and \quad $ r\longrightarrow\lambda r$.
 These metrics are exact solutions to gravitational theories coupled to an appropriate matter,
with an abelian gauge field in the bulk. By including an abelian gauge field and scalar
dilaton, one can construct the full class of metrics \cite{ppc,hoc,ptb,hcd,eff,dwh,tbc,hddb,hfn,cdab,hfs,hfsi,heef}, so which is given by,
\begin{equation}\label{metric2}
ds^2_{d+2}=r^{-2\frac{d-\theta}{d}} (-r^{-2(z-1)} dt^2+dr^2+dx_{i}^2),
\end{equation}
where $z$ and  $\theta$  are  dynamical critical exponent of hyperscaling violation metric  \cite{scsd}. As mentioned before, this metric is not scale invariant under above scaling.
So, if  we consider  some finite temperature in this theory we have to account some $f(r)$  in  the corresponding metric which is given by the following equation,
\begin{equation}\label{metric3}
ds^{2}_{d+2}=e^{2A(r)} (-e^{2B(r)} f(r) dt^2+ \frac{dr^2}{f(r)}+dx_{i}^2).
\end{equation}
Then the temperature is proportional to a power of $r_{h}$ and for general $A(r)$ and $B(r)$ the temperature also depends on them. Moreover, in the gravity side we should take $r_{F} <r_{h}$ and $r_F$ is the inverse scale of the Fermi surface. In order to have black hole solution we have to consider following background \cite{hfnf},
\begin{equation}\label{metric4}
ds^{2}_{d+2}=\frac{R^2}{r^2} (\frac{r}{r_{F}})^{\frac{2\theta}{d}} (-r^{-2(z-1)}f(r) dt^2+\frac{dr^2}{f(r)}+dx_{i}^{2}),
\end{equation}
with $f(r)=1-(\frac{r}{r_{h}})^{d+z-\theta}$ , $r_{F}<r_{h}$ and the temperature will be as $T=\frac{1}{4\pi} \frac{\vert d+z-\theta \vert}{r^{z}_{h}} .$
Here we note that the  the Null Energy condition imposes $(d-\theta)(d(z-1)-\theta)\geq 0$, \quad \quad $(z-1)(d+z-\theta)\geq 0$ \cite{ucbe}.\\
\subsection{Review on results of thermal width of a meson in hyperscaling violation background}
One important constraint on thermal width is negativity of $ImV_{Q\bar{Q}}$. It means that In QCD system the imaginary part of potential shows the decay behaviour and the corresponding potential should be negative. Keeping this condition in mind, we found imaginary part of potential for a moving meson in plasma by holography approach with a parametric HSV background. On the other hand there are some string solutions for such background could be used in a gravity side of gauge/gravity duality approach. By checking them  to know if they satisfy  $ImV_{Q\bar{Q}}<0$, we  found two special cases of the hyperscaling violation metric background parameters as, $z=1$, $\theta=4$, $d=1 or 2$ and $z=0$, $\theta=3$, $d=1 or 2$ where both categories satisfy the mentioned condition. We have shown  that hyperscaling violating metric is very close to a YM theory  at finite temperature with suitable $d$, $z$ and $\theta$. Thereafter, the thermal width of a moving meson in a plasma was shown  by this metric. Briefly, this is what has been done in \cite{hsvinve}. We have some interesting results which can be compared with entropic force, since both thermal width and entropic force are responsible for thermal dissociation.

\section{Entropic force of a moving dipole in HSV background }

We start with HSV metric background and derive the entropic force relation for a meson which is moving perpendicular in a plasma with rapidity $\xi$. Then we apply it for $z=1$, $\theta=4$ and $d=2$ case.
\subsection{ Perpendicular case }
 We have a meson with velocity perpendicular to the wind and
rapidity $\xi$ in the plasma then we can assume the pair is at rest and the frame is moving in reverse direction with $-\xi$ \cite{fn}. We continue with hamiltonian to find inter-quark distance
length as,
\begin{equation}\label{eq:hamil}
H(r)\equiv\sqrt{\frac{\tilde{V}(r)}{\tilde{V}_{c}}\frac{\tilde{V}
(r)-\tilde{V}_{c}}{\tilde{M}(r)}},
\end{equation}
with this definition,
\begin{eqnarray}\label{mtilda}
\tilde{M}(r)\equiv g_{00}g_{rr}\cosh ^2 \xi -g_{xx}g_{rr}\sinh ^2 \xi\ ,
\end{eqnarray}
\begin{eqnarray}\label{vtilda}
\tilde{V}(r)\equiv g_{00}g_{xx}\cosh ^2 \xi -{g_{xx}}^2\sinh ^2 \xi\ .
\end{eqnarray}
Where $g_{00}$, $g_{xx}$ and $g_{rr}$ are components of the corresponding metric and $\tilde{V_c}$ means $\tilde{ V (r_c)}$ and c is the deepest position of the string in the bulk.\\ 
The equation of motion and the boundary conditions of the string relates $L$ (length of
the line joining both quarks) with $r_c$ as follows,
\begin{equation}\label{eq:lstr}
\frac{L}{2}=-\int_{r_{c}}^{\infty}\frac{dr}{H(r)}.
\end{equation}
As we mentioned before, the entropic force has been defined as the entropy $S$ which grows with the inter-quark distance $L$. So we have entropic force as,
\begin{equation} \label{entfor}
\mathcal{F}=T\frac{\partial S}{\partial L},
\end{equation}

where $T$ refers to temperature of the plasma.
Then one can calculate the entropy as,
\begin{equation} \label{eq:entrop}
S=-\frac{\partial F}{\partial T}.
\end{equation}

Now there are two suppositions in here, first for large values of inter-quark distance, where it becomes larger than the maximum value of $LT$ (which we call $c^{\prime}$), therefore the quarks are compeletly screened because the fundamental string is breaking in two pieces. The second one means that the fundamental string is connected yet and $LT$ is smaller than $c^{\prime}$.\\
 If $L>\frac{c'}{T}$ free energy is not unique and it depends on configuration of strings, so from \cite{desc} we choose it as,
\begin{equation} \label{eq:F1}
F^{(1)}=\frac{1}{\pi\alpha'}\int_{r_h}^{\infty} dr.
\end{equation}
According to (\ref{eq:entrop}), we get numeric result as,
\begin{equation} \label{eq:S1}
S^{(1)}=\sqrt{\lambda}\Theta(L-\frac{c'}{T}),
\end{equation}

where $\Theta$ is mathematical theta function, and the entropy of moving quarkonium
for large distance of $L$, is similar to the static case which have been considered in many references \cite{edm, tech} so we skip from this step and proceed by the second case. \\
If: $L<\frac{c'}{T}$, the free energy which is derived from the on-shell action of the fundamental string in the dual geometry is as follows,
\begin{eqnarray} \label{eq:fre}
F^{(2)}&=&\frac{1}{\pi\alpha'}\int_{r_{c}}^{\infty}dr\sqrt{\frac{\tilde{M(r)}\tilde{V(r)}}{\tilde{V(r)}-\tilde{V_{c}}}}\nonumber\\
\end{eqnarray}
\subsection*{Considering $z=1$, $\theta=4$ and $d=2$ }
\begin{figure}[h!]
\begin{center}$
\begin{array}{cccc}
\includegraphics[width=100 mm]{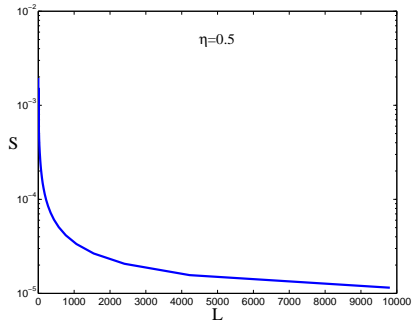}
\end{array}$
\end{center}
\caption{Entropic force of a moving meson in our condidate  HSV background, with small rapidity and perpendicular velocity vector to the joining axis of the pair.}
\label{fig1}
\end{figure}

By considering $z=1$, $\theta=4$ and $d=2$, which behave  very close to YM theory\cite{hsvinve}, we will find entropic force. In addition, note that $\xi$ should be very small \cite{hsvinve}. From  (\ref{eq:lstr}), (\ref{entfor}), (\ref{eq:entrop}) and (\ref {eq:fre}), we follow our studying.\\
 In figure (\ref{fig1}) the entropic force of a moving meson in plasma has been considered. We have shown that in small rapidity according to the condition  obtained in \cite{hsvinve}. One can see the entropic force has a decreasing behaviour corresponds to increasing thermal width and increasing dissociation length of the pair. These results are in agreement with each other.
\subsection{Arbitrary angles}
In this section we extend our calculations for arbitrary angles, it means that orientation of the
dipole can have any arbitrary angle with respect to the velocity vector. $\alpha$ is the angle of the dipole with
respect to the $X_{d-1}$ and the dipole is on the $(X_1,X_{d-1})$ plane.  From the boundary conditions and the action, constants of motion can be found. Proceeding by them  the following relation will result,
\begin{align}\label{eq:lcos}
\frac{L}{2} \cos\alpha =- K \int_{r_c}^{\Lambda}\, dr \, \sqrt{\frac{\left[M(r) \cosh^2 \xi - N(r)  \sinh^2 \xi \right] \left[V(r) \cosh^2 \xi - P(r)  \sinh^2 \xi\right]}{V(r)\left\{(V(r)-K^2)\left[V(r) \cosh^2 \xi - P(r)  \sinh^2 \xi\right] - V(r) q^2 \right\}}}.
\end{align}
It is worth  noting that the exact derivation of above calculations can be found in \cite{sif} in detail.
Now, we are going to extend the previous discussion about the entropic force when meson is moving with the velocity which has arbitrary angle with respect to the wind. According to what we mentioned before, there are two suppositions: the first one when quarks are compeletly screened and we have a large value of inter- quark distance length. In that case according to (\ref{eq:F1}) the free energy does not depend on the angle of velocity with respect to the wind. Therefore the corresponding entropic force will be same  as last section. But when fundamental string is connected yet and $LT$ is smaller than its maximum value, the free energy is,
\begin{align}\label{entforarb}
F^{(2)} & =- \frac{\mathcal{T}}{\pi \alpha'} \int_{r_c}^{\Lambda} \, dr \, \left\{ \sqrt{\frac{ V(r) \left[M(r) \cosh^2 \xi - N(r)  \sinh^2 \xi \right] \left[V(r) \cosh^2 \xi - P(r)  \sinh^2 \xi\right]}{ \left\{(V(r)-K^2)\left[V(r) \cosh^2 \xi - P(r)  \sinh^2 \xi\right] - V(r) q^2 \right\}}} \right.  
\end{align}
\subsection*{Considering $z=1$, $\theta=4$ and $d=2$ in arbitrary angle case}
\begin{figure}[h!]
\begin{center}$
\begin{array}{cccc}
\includegraphics[width=100 mm]{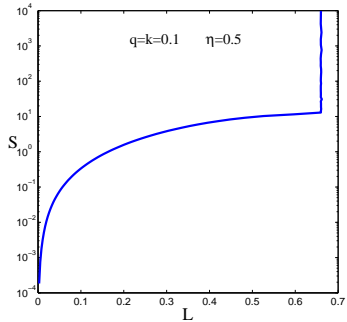}
\end{array}$
\end{center}
\caption{Entropic force of a moving meson in our condidate  HSV background, with small rapidity and arbitrary angle of velocity vector with joining axis of the pair.}
\label{fig2}
\end{figure}

By considering $z=1$, $\theta=4$ and $d=2$, we extend our studying for arbitrary angle of the pair with wind. Again note that $\xi$ should be very small \cite{hsvinve}. From   (\ref{eq:lcos}) and (\ref {entforarb}), we follow our studying. In figure (\ref{fig2}) the entropic force  of a moving meson in plasma has been considered. We show it in small rapidity according to the condition  obtained in \cite{hsvinve}. One can see the entropic force has a decreasing behaviour corresponds to increasing thermal width and increasing dissociation length of the pair. These results are in agreement with each other.
\section{Diffusion constant in a HSV background}
Entropy density is related to shear viscosity $\eta$, by the relation $\frac{\eta}{s}=\frac{1}{4\pi}$ for a wide variety of theories. By using holographic approach in HSV gravity duals, it is interesting to study hydrodynamic behaviour showed by diffusion constant.\\
There are another category of the exact values of hyperscaling violation background as $d,z,\theta=2,0,3$ found by condition $ImV_{Q\bar{Q}}$ in \cite{hsvinve}, implys that there are such parameters in Lifshitz-like theories lead to a YM-like theory which satisfy thermal width condition for a slowly moving meson in a plasma  in limit $r_c\longrightarrow r_h$. As we used the first category in last sections to find entropic force in a YM like theory with HSV background, we will study diffusion constant with exact relation found in \cite{hypsh} for $d-z-\theta=-1$.
\subsection{Diffusion constant in case $d-z-\theta=-1$}
The main idea is to use approach of \cite{hypsh} to consider the diffusion constatnt in HSV backgrounds. From there in case hyperscaling violation background parameters satisfy $d-z-\theta=-1$ one finds difussion constant as, 
\begin{equation}
\label{diffhsv}
\mathcal{D}=r_0^{d-\theta-1}log(\frac{r_h}{r_c})=r_0^{z-2} log(\frac{r_h}{r_c}).
\end{equation}
 So, with $d,z,\theta=2,0,3$ and in limit $r_c\longrightarrow r_h$, we arrive at,
 \begin{equation}
\label{diffhsv2}
\mathcal{D}=0,
\end{equation}
 which shows there is hydrodynamical equilibrium in system.\\
  So breifly, in a YM-like theory described by a HSV gravity dual, system of a very slowly moving meson  in a plasma is in equilibrium.
\section{Conclusion}
We had a comparison between enropic force and thermal width of a moving pair in plasma, since both of them are related to meson dissociation. In \cite{hsvinve} the satisfying parameters of hyperscaling violation metric background from view point of condition  $ImV_{Q\bar{Q}}<0$ have been introduced. In addition from there we know metric with  $z=1$, $\theta=4$ and $d=2$ is in agreement with above condition besides such HSV metric behave similar to YM theory. So proceeding by this choice we found entropic force of a slowly moving quarkonia in a plasma.\\
 Entropic force and thermal width are related to thermal dissociation both, therefore studying of them could have interesting results. With this motivation and all above considerations we calculated entropic force in two cases, when velocity vector of meson is perpendicular to joining axis of the pair and then we extended those calculation to arbitrary angles. We considered that entropic force has a decreasing behaviour with length which corresponds to increasing dissociation length. These results are in agreement with \cite{hsvinve} which presented suitable HSV candidate   similar  to YM theory. In addition such  system has zero shear diffusion canstant which describes an equilibrium.\\

\section*{\textbf{Acknowledgement}}
The authors are grateful very much to Fatemeh Razavi for support and valuable activity in
numerical calculations and also Zeinab Amoozad for useful discussion.

\end{document}